\documentclass[11pt,preprint]{revtex4}  
\usepackage{amsmath}   
\usepackage{amssymb}   
\usepackage[T1]{fontenc} 
\usepackage{graphicx}  

\begin{document}
\title{Droplet-shaped waves: causal finite-support analogs of X-shaped waves}

\author{Andrei B Utkin}
\affiliation{INOV - Inesc Inova\c{c}\~{a}o and 
ICEMS, Instituto Superior T\'{e}cnico, 
Technical University of Lisbon, Av. Rovisco Pais, Lisbon 1049-001, Portugal}
\email{andrei.utkin@inov.pt}

\begin{abstract}
A model of steady-state X-shaped wave generation by a superluminal (supersonic) point-like source infinitely moving along a straight line is extended to a more realistic causal scenario of a source pulse launched at time zero and propagating rectilinearly at constant superluminal speed. In the case of infinitely short (delta) pulse, the new model yields an analytical solution, corresponding to the propagation-invariant X-shaped wave clipped by a droplet-shaped support, which perpetually expands along the propagation and transversal directions, thus tending the droplet-shaped wave to the X-shaped one.


\end{abstract}

\maketitle
\section{Introduction}
\label{sec:1}

The early 1970s witnessed the start of the intensive study of radiation emanated 
by superluminal sources 
\cite{Gron1978,Lemke1975,RecamiBook1978,RecamiReview_RivistaNuovoCim_1986},
which received a new impetus due to possibility of launching propagation-invariant 
localized waves 
\cite{LocWaves,RecamiEtAl_LocSupLumSolME_2003,Abdollahpour2010,Arevalo2010,SaariReivelt1997}, 
whose fundamental representations and classification schemes are discussed in 
\cite{BesierisEtAl_1998,SaariReivelt_2004}. Among different types of the localized
wave structures, the X-shaped waves are one of the most well-known 
\cite{SaloEtAl_2000,DiTrapani2003,PorrasDiTrapani2004}. 
Recami et al. \cite{RecamiZamboniRachedDartora2004} proposed a toy model of localized 
X-shaped field generation by an axisymmetric superluminal four-current 
${{J}^{\left( \alpha  \right)}}=\left( c\varrho ,0,0,j \right)$, 
where $c$ is the speed of light, $\varrho $ the electric charge density, and $j$ the 
only non-zero component of the current density 
vector $\mathbf{j}$ in the cylindrical coordinates $\rho ,\varphi ,z$. Considering 
the inhomogeneous electromagnetic problem for the four-potential in the form
${{A}^{\left( \alpha  \right)}} =\left( A^{\left( 0 \right)},\mathbf{A} \right)
  =\left( A^{\left( 0 \right)},0,0,\psi \right)$, 
${{A}^{\left( \alpha  \right)}}\left( \tau ,\rho ,\varphi ,z \right)
={{A}^{\left( \alpha  \right)}}\left( \rho ,\zeta  \right)$, $\alpha =0,1,2,3$,
where $\zeta =z-Vt=z-\beta \tau $ is a $V$-cone variable, introduced for constant 
superluminal velocity $V=c\beta$, and $\tau =ct$ the time variable, the investigation 
was inevitably constrained to the cases that
(i) describe a steady-state process that lasts all times from $-\infty $ to $\infty $,
(ii) require similar four-current dependence ${{J}^{\left( \alpha  \right)}}\left( \tau, \rho ,\varphi ,z \right)={{J}^{\left( \alpha  \right)}}\left( \rho ,\zeta  \right)$, and
(iii) bound the nonzero components of the four-vectors by relations 
$j\left( \rho ,\zeta  \right)= c \beta {\varrho }\left( \rho ,\zeta  \right)$ 
and ${\psi}(\rho ,\zeta )=\beta A^{\left( 0 \right)} (\rho ,\zeta )$ due to, 
respectively, the continuity equation and the Lorentz gauge. 

The problem was solved for the case of the rectilinear motion of a point-like charge 
using the spectral method (Fourier-Bessel and Fourier transforms with respect to 
$\rho $ and $\zeta $), yielding a localized "true X wave" solution
\begin{equation} \label{RecamiPotentialA} 
{\psi}\left(\rho, \zeta \right) =const\times \frac{\beta }{\sqrt{{{\zeta }^{2}}-\left( {{\beta }^{2}}-1 \right){{\rho }^{2}}}}.
\end{equation}
Being symmetric with respect to time reversal and involving equally the advanced and 
retarded components, such a model of X wave generation is acausal. Later attempts to 
remedy this drawback introducing the unilateral conical-wave source 
\cite{WalkerKuperman2007} and applying the unidirectional decomposition 
\cite{ZamboniRachedPRA2009,ZamboniRachedJOSA2010} resulted in more realistic models.
All these studies, however, do not take into account the temporal constraint of the source 
motion -- even in its weak form that assumes the existence of an initial moment, 
prior to which neither the source pulse nor the emanated wave can exist.

\section{Statement of the problem}
\label{sec:2}

The present work considers wave generation by superluminal charges within the framework
of the Riemann-Volterra approach, introducing a direct space-time domain method that, 
in contrast to constructing the wavefunctions via the Bateman transform 
\cite{Besieris2004,UtkinComTeorPhys2011} or Green's method \cite{ZamboniRachedJOSA2010}, 
deals with the initial value problem that includes, in addition to the inhomogeneous
wave equation 
\begin{equation} \label{InhomogenWaveEq} 
\left[ {\partial _\tau ^2 - {\rho ^{ - 1}}{\partial _\rho }\left( {\rho {\partial _\rho }} \right) - \partial _z^2} \right]\psi \left( \tau, \rho ,z \right) = S \left( \tau, \rho ,z \right) ,
\end{equation}
the time asymmetric initial condition
\begin{equation} \label{TimeAsymIniCond} 
\psi  = 0,\;S = 0\quad {\rm{for}}\quad \tau  < 0,
\end{equation}
which defines the {\it arrow of time} and, as will be shown, represents a sufficient condition for
existence of a unique causal solution. 
For the electromagnetic waves, we assume that $\psi$ and $S = \mu_0 j$ represent, respectively, the $z$-components of the four-potential ${A^{(\alpha )}}$ and four-current ${J^{(\alpha )}}$ ($\mu_0$ stands for the magnetic constant). Furthermore, let us get over one singularity, considering, instead of the line $\delta$-pulse, a line source of arbitrary shape $f\left( \xi  \right) = f\left( -\zeta  \right) = f\left( \beta\tau - z \right) $ propagating with a constant superluminal speed in the positive $z$-direction. In view of (\ref{InhomogenWaveEq}), (\ref{TimeAsymIniCond}), such a source can be expressed in the form
\begin{equation} \label{SourceCurrent} 
j\left( \tau, \rho ,z \right) = ce \beta \frac{\delta \left( \rho  \right)}{2\pi \rho }
  h\left( z \right)h\left( \beta \tau -z \right)f\left( \beta \tau -z \right),
\end{equation}
%
where $e$ is the elementary charge while $\delta(\cdot)$ and $h(\cdot)$ stand for 
the Dirac delta and Heaviside step functions. 

One can easily notice that due to (\ref{TimeAsymIniCond}) in no case the problem in question 
can yield a pure localized solution meeting the propagation-invariant condition stated by 
Lu (Ref. \cite{LocWaves}, p.~98) as reducibility to the form 
$\psi \left( \tau, \rho ,z \right) = \psi \left(\rho , \xi \right)$.
Instead, it provides us with bounded-support, finite-energy wavefunctions (signals) that 
can serve as real-world, physically admissible approximations of the ideal temporally 
unbounded localized waves. Borrowing the terminology from the numerical analysis, we 
may say that the description of a localized propagation-invariant wave by such a signal 
is convergent if the signal asymptotically approaches the localized wave structure as 
$\tau \rightarrow \infty$.

The continuity equation $c \partial_\tau \varrho + \partial_z j = 0$ and
representation (\ref{SourceCurrent}) immediately implies that the charge 
density is uniquely defined by equation
\begin{equation} \label{FormulaForElCharge} 
\begin{split}
&\varrho\left( \tau, \rho ,z \right) = -\frac{1}{c} \int_0^\tau{d {\tau }'\partial_z j}
 = e \frac{\delta \left( \rho  \right)}{2\pi \rho }
              \left[ \vphantom{\int_{0}^{\beta \tau }}
             h\left( z \right)h\left( \beta \tau -z \right)
             f\left( \beta \tau -z \right) 
           \right. \\
        &- \left.
             \beta h\left( \tau \right) \delta \left( z \right)
             \int_0^\tau{d {\tau }'f\left( \beta {\tau }'-z \right)} 
           \right] = \frac{1}{c\beta }j+
           \left( -e \right)\frac{\delta \left( \rho  \right)}{2\pi \rho }h\left( \tau  \right)\delta \left( z
           \right)
           \int_0^{\beta \tau}{d \xi f \left( \xi \right)},
\end{split}
\end{equation}
%
and the other non-zero component of the four-potential ${{A}^{(0)}}$ -- automatically satisfying 
$\square {{A}^{(0)}}={{\mu }_{0}}{{J}^{(0)}}$ -- can be obtained as a consequence of 
(\ref{InhomogenWaveEq}), (\ref{TimeAsymIniCond}), and (\ref{FormulaForElCharge}) 
from the Lorenz gauge
$\sum\nolimits_{\alpha }{{\partial_\alpha}{{A}^{(\alpha )}}=0}$ by a simple integration
\begin{equation} \label{FormulaForAo} 
A^{(0)} = -\int_{0}^{\tau }{d \tau' \partial_z A^{(3)}}
        = -\int_{0}^{\tau }{d \tau' \partial_z \psi}.
\end{equation}
Here the second term describes accumulation of the opposite charge at the pulse-generation point 
$\rho =0,z=0$ due to current outflow. 

For models concerning the wave emanation by hypothetical 
superluminal particles, tachyons, such an accumulation can be nulled if we suppose that the 
generation of a particle is accompanied by generation of an antiparticle (an opposite point-like 
charge) moving in different direction.

For other models involving more traditional source structures the excessive charge 
accumulation can be avoided in the long term by considering (bipolar) current pulses for which 
$\int_{0}^{\infty }{d \xi f\left( \xi  \right)}$ is vanishing or, at least, limited (demonstrating,
for example, oscillating behavior).
One of such embodiments discussed by Ziolkovski et al. \cite{ZiolkowskiEtAl_1993} and 
later by Saari \cite{Saari_2001} envolves superluminal sink-source-type fictitious 
currents \cite{SheppardSaghafi_1998}. 
As discussed in \cite{Saari_2001} with reference to Fig.~1 therein, 
the simplest and most straightforward implementation of the superluminal wave motion can be 
achieved with a type of "scissor effect", creating an interference pattern of two 
(luminal-speed) plane waves whose fronts are inclined with respect to each other. Similar patterns
can be created, e.g., at oblique incidence of wavefronts onto locally plane targets (screens).
On the cosmic scale, this phenomena is well discussed by astrophysicists (see, e.g., 
\cite{AstrophJets_1993}). Remarkably, recent advances in ultraintense laser interaction 
science enables such huge-scale phenomena to be reproduced in the micrometer scale 
in the laboratory conditions.
Excitation of a source current pulse by an ultra-intense, ultra-short pulsed laser beam, 
incident at an angle to a thin plasma filament --- 
induced in a medium by another laser beam --- is an example (here significant current
density can be obtained due the effects akin to the wake-field acceleration while 
the desired orientation of the macroscopic current along the filament can be asserted 
by imposing a strong magnetic field; description of pertinent phenomena can be found 
in \cite{SilvaLuisO_1995}).

\section{General solution}
\label{sec:3}

\begin{figure}[h]
\begin{center}
\includegraphics*[width=0.5 \textwidth]{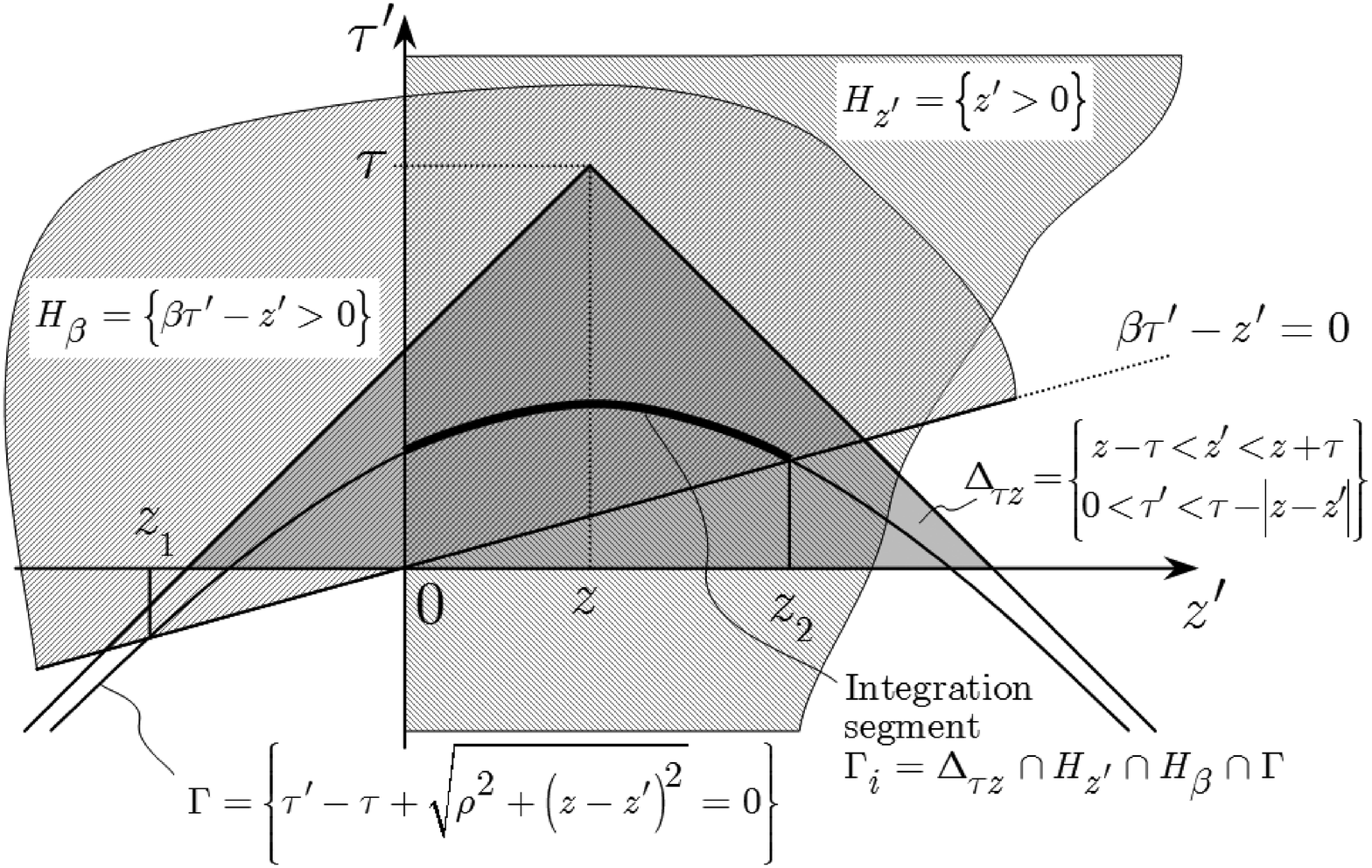}
\end{center}
\caption{Integration path for general solution (\ref{DblIntegralSolution}).}
\label{Fig_integration_segment}
\end{figure}

Separating $\rho $ by the Fourier-Bessel transform
$
\Psi\left( \tau, s, z \right)
=\int_{0}^{\infty }d \rho\,\rho{\psi\left( \tau, \rho, z \right){{J}_{0}}\left( s\rho \right) },
$
one gets from (\ref{InhomogenWaveEq}), (\ref{TimeAsymIniCond}) and (\ref{SourceCurrent}) a simpler
time asymmetric initial value problem for the Klein-Gordon equation
\begin{equation} \label{KGE} 
\left( \partial^2_\tau - \partial^2_z + s^2 \right) \Psi
= \frac{{{\mu }_{0}}}{2\pi }
ce\beta 
h\left( z \right)  h\left( \beta \tau -z \right)  f\left( \beta \tau -z \right),
\end{equation}
with the initial condition 
\begin{equation} \label{ConditionForKGE} 
\Psi = 0 \quad \rm{for} \quad \tau  < 0.
\end{equation}
One can easily check by direct enumeration that condition (\ref{ConditionForKGE}) uniquely defines 
the transition to the (characteristic) variables 
$z,\tau \rightarrow {1 \over 2}(\tau + z), {1 \over 2}(\tau - z)$
that: (i)~transforms the Klein-Gordon equation to the first canonical form and 
(ii)~results in the geometry of the solution support, the area of $\Psi \neq 0$, 
corresponding to the well-established layout of the Riemann-Volterra method 
(see, e.g., Fig.~30 in \cite{CourantHilbert1989} or Fig.~1 in \cite{Smirnov1964}). 
The {\it ad hoc} Riemann-Volterra procedure involving known Riemann function 
${{J}_{0}}\left( s\sqrt{{{\left( \tau -{\tau }' \right)}^{2}}-{{\left( z-{z}' \right)}^{2}}} \right)$ 
(see \cite{UtkinBookCh2011} for details) readily provides a unique causal solution 
\cite{Smirnov1964}, which after the inverse transition to the independent variables 
$z, \tau$ takes the form
\[
\begin{split}
 \psi \left(\tau, s, z \right) &= \frac{\mu_0}{4\pi } c e \beta
  \int_{z-\tau }^{z+\tau }{
      d z'\int_{0}^{\tau -\left| z-{z}' \right|}{
        d \tau' {{J}_{0}}\left( s\sqrt{{{\left( \tau -{\tau }'
    \right)}^{2}}-{{\left( z-{z}' \right)}^{2}}} \right)\,}} \\ 
 &\times \ h\left( {{z}'} \right)h\left( \beta {\tau }'-{z}' \right)
    f\left( \beta {\tau }'-{z}' \right).
\end{split}
\]
Using the inverse Fourier-Bessel transform, the closure equation (see, e.g., \cite{Arfken2001}, p. 691)
\[
\int_0^\infty{d s\,s J_0 \left( s \rho \right) J_0 \left( s \rho' \right)}
  = \frac{1}{\rho} \delta \left( \rho - \rho' \right)
\]
and the representation of the delta function with simple zeros $\lbrace \tau_i \rbrace$ (\cite{Arfken2001}, p. 87)
\begin{equation} \label{DeltaFuncWithSimple0}
\delta \left( g \left( \tau \right) \right) =
\sum_{i}{
  \frac
    {\delta \left( \tau-{{\tau}_{i}} \right)}
    {\left| \partial_\tau g \left( \tau_i \right) \right|}}
\end{equation}
%
(among the two simple zeros, only one lies within the integration limits), we get the solution to the initial space-time domain problem (\ref{InhomogenWaveEq}), (\ref{TimeAsymIniCond}) in the form
\begin{equation} \label{DblIntegralSolution}
\begin{split}
\psi \left(\tau, \rho, z \right) &= \frac{\mu_0} {4\pi \rho } ce\beta
  \int_{z-\tau }^{z+\tau }{
    d z' \int_0^{\tau -\left| z-{z}' \right|}{  
      d \tau' h \left( z' \right) 
    }
  } \\
&\times h \left( \beta \tau' - z' \right) f \left( \beta \tau' - z' \right)
    \frac
     {\delta \left( {\tau }'-\tau +\sqrt{{{\rho }^{2}}+{{\left( z-{z}' \right)}^{2}}} \right)} 
     {\sqrt{\rho^2 + \left( z - z' \right) ^2 } } .
\end{split}
\end{equation}
%
As illustrated in Fig.~\ref{Fig_integration_segment}, $\psi$ is defined by integration over the curve segment 
$\Gamma_i = \Delta_{\tau z} \cap H_{z'} \cap H_\beta \cap \Gamma$
that is the intersection of: the initial triangle integration area 
${{\Delta }_{\tau z}}=
\{z-\tau <{z}'<z+\tau ,\ 0<{\tau }'<\tau -\left| z-{z}' \right|\}$; 
the half-plane 
${{H}_{\beta }}=\left\{ \beta {\tau }'-{z}'>0 \right\}$, 
the support of the step function 
$h\left( \beta {\tau }'-{z}' \right)$; 
the half-plane ${{H}_{{{z}'}}}=\left\{ {z}'>0 \right\}$, 
the support of the step function $h\left( {{z}'} \right)$; 
and the hyperbola 
$\Gamma =\{ {\tau }'-\tau +\sqrt{{{\rho }^{2}}+{{\left( z-{z}' \right)}^{2}}}=0 \}$, 
the support of the delta function 
$\delta ( {\tau }'-\tau +\sqrt{{{\rho }^{2}}+{{\left( z-{z}' \right)}^{2}}}\:)$.
%
%
%

\begin{figure}[h]
\begin{center}
\includegraphics*[width=0.5 \textwidth]{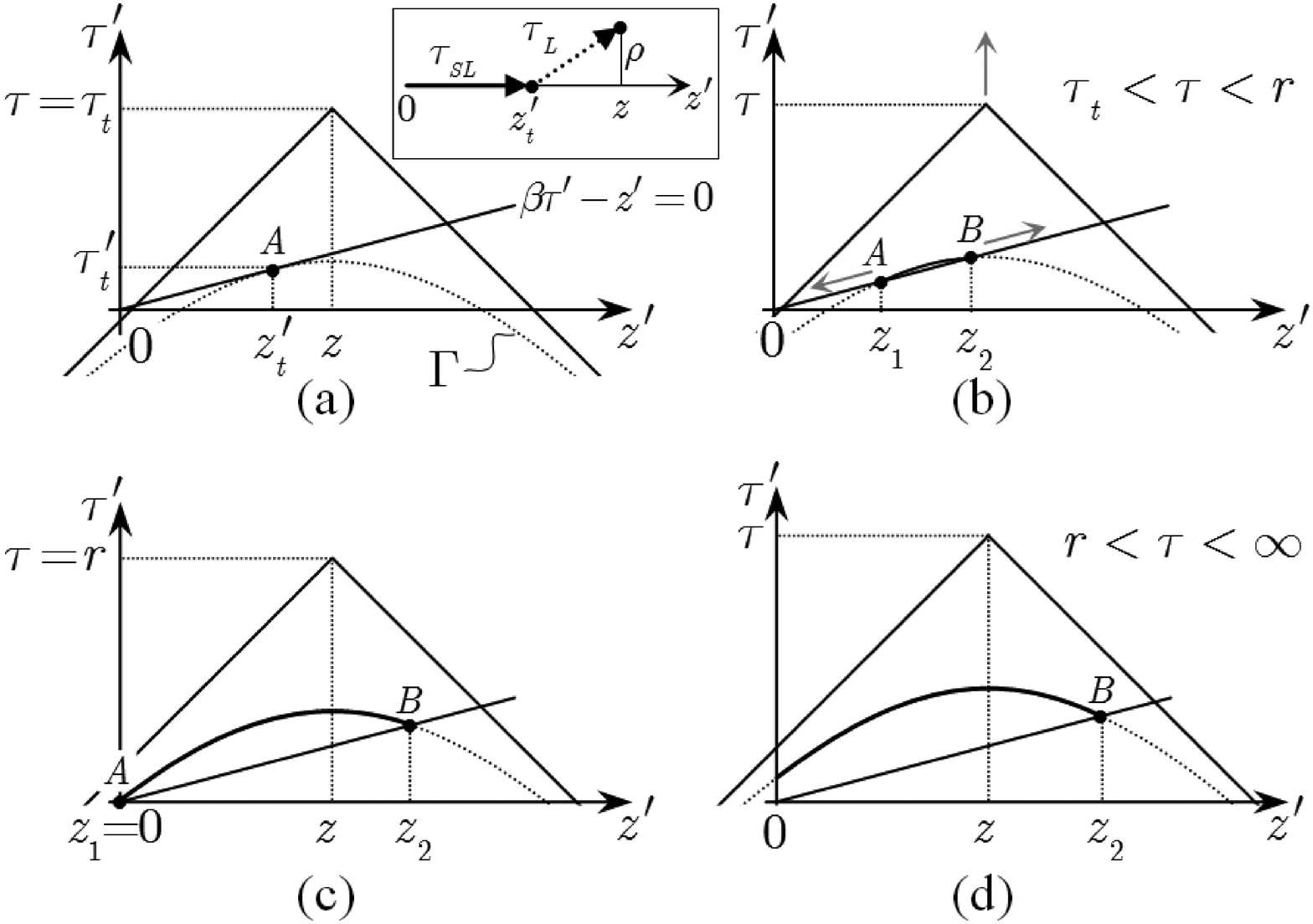}
\end{center}
\caption{Instantiation of general solution for the observation point $\rho, z$, case $z > z_c$.}
\label{Fig_zp_taup_sequence}
\end{figure}

Instantiation of general formula (\ref{DblIntegralSolution}) with the help of the 
${z}',{\tau }'$ plane diagrams 
(Fig.~\ref{Fig_zp_taup_sequence}) reveals the explicit structure of the solution,
%
%
%

if $-\infty < z < z_c  \; \Leftrightarrow \; \tau_t > r$,
\begin{equation} \label{PsiOrdinaryCases}
\psi  =  \left\{ 
\begin{array}{ll}
   0 & \; -\infty <\tau < r  \\
   \frac{{{\mu }_{0}}}{4\pi }ce\beta 
     \int_0^{z_2}{d z' \frac{f \left( \beta \left( \tau - r' \right) - z' \right)}{r'}} & 
   \; r < \tau < \infty
\end{array} 
\right.
\end{equation}
%

otherwise ($z_c < z < \infty \; \Leftrightarrow \; \tau_t < r$)
\begin{equation} \label{PsiExtraordinaryCases}
\psi  =  \left\{ 
\begin{array}{ll}
   0 & \; -\infty < \tau < \tau_t \\
   \frac{{{\mu }_{0}}}{4\pi }ce\beta 
     \int_{z_1}^{z_2}{d z' \frac{f\left( \beta \left( \tau -{r}' \right)-{z}' \right)}{r'}} & 
   \; \tau_t < \tau < r  \\
   \frac{{{\mu }_{0}}}{4\pi }ce\beta 
     \int_{0}^{{{z}_{2}}}{d z' \frac{f\left( \beta \left( \tau -{r}' \right)-{z}' \right)}{r'}} & 
   \; r < \tau <\infty
\end{array}
\right.
\end{equation}
%
that depends on the following parameters:
$ {{z}_{1,2}} =\beta \gamma^2
  \left[ \beta z - \tau \mp \sqrt{{{\left( \beta \tau -z \right)}^{2}} -
  \left( {{\beta }^{2}} - 1\right){{\rho }^{2}}} \right]$,
$z_c = \gamma \rho$, $\gamma =(\beta^2-1)^{-1/2}$,
$\tau_t =\left( z+\rho /\gamma \right)/\beta$, 
$r = \sqrt{{{\rho }^{2}}+{{z}^{2}}}$, and 
$r'  = \sqrt{{{\rho }^{2}}+{{\left( z-{z}' \right)}^{2}}}$.
%
The result illustrates capability of the method -- that does not resort to \textit{a priory} 
causality conditions like retarded nature of the argument, wave propagation in a fixed direction, etc. 
-- to yield solutions whose space-time structure admits easy \textit{posterior} interpretation 
in terms of causal propagation of information:

For an observation point $\rho, z$ there is no intersection between $H_\beta$ and $\Gamma $ 
(no wave registered) until the moment $\tau = \tau_t$, 
in which the hyperbola touches the boundary of $H_\beta$ in the (tangency) point 
$\tau'_t =\left( z-\gamma \rho  \right)/\beta $, $z'_t = z - \gamma \rho $, as depicted in  Fig.~\ref{Fig_zp_taup_sequence}(a).
For $z > z_c$ the value of $\tau_t = \tau_{SL} + \tau_L$ represents the minimum time necessary for the electromagnetic energy, originated in the space-time point 
$\tau' = \rho' = z' =0$, to reach the observation point $\rho'=\rho $, $z'=z$ -- first propagating during time 
$\tau_{SL} = \tau'_t =\left( z-\gamma \rho  \right) / \beta $ along the $z'$ axis at the superluminal speed $c \beta$ with the source-pulse front and then during time 
$\tau_L =\sqrt{\rho^2 + \left( z- z'_t \right)^2} = \beta \gamma \rho$ 
out of this axis toward $\rho ,z$ at the luminal speed with the front of the emanated electromagnetic wave
(see the upper diagram of Fig.~\ref{Fig_zp_taup_sequence}(a)).

For cases corresponding to Figs.~\ref{Fig_zp_taup_sequence}(a,b), Eq.~(\ref{PsiExtraordinaryCases}) 
characterizes quantitatively a source pulse that appears to the observer 
"suddenly growing out of a point" $z'_t, \tau'_t$ at $\tau = \tau_t$ and 
expanding in the opposite directions, just as was predicted by Gron \cite{Gron1978}
on the basis of purely geometrical consideration 
(see also Fig. 15 of \cite{RecamiReview_RivistaNuovoCim_1986} or 
Fig. 1 of \cite{Barut_1982} ). 
Notably, apart from \cite{Gron1978},
other pioneering papers showed how a superluminal source, after having suddenly 
revealed in an "optical-boom" phase, may subsequently appear as a couple of objects 
receding one from the other, in particular, in astrophysical observations (so-called
"superluminal expansions"). For a detailed discussion of these models, both "orthodox" 
--- in which a superluminal pattern is created by a coordinated motion of the subluminally 
moving constituents --- and "true superliminal" --- in which actual superluminal motions are 
treated within the framework of the extended special relativity theory ---, the interested reader 
is directed to Recami et al. \cite{RecamiEtAl_SpLumExpan_1986} and references therein.

Causal solution (\ref{PsiExtraordinaryCases}) demonstrates other phenomena discussed 
in the literature from the standpoint of the extended special relativity: as expound in 
\cite{RecamiReview_RivistaNuovoCim_1986}, the superluminal (tachyonic) motions are always observed
being forward in time, but can appear reversed in direction; as well, the duality between 
the source and detector makes possible the emission to be observed as absorption. 
In the case in question, the "post-boom" evolution of the source pulse is observed as forward
and backward expansion. The latter stops as the pulse reaches its origin $z'=0$
(Figs.~\ref{Fig_zp_taup_sequence}(c,d)) and at this point the (apparently reversed) 
generation process manifests as the pulse absorption. 
In the limiting case of a $\delta$-pulse the half-plane $H_\beta$ degenerates into a 
line and the segment of integration $\Gamma_i$ into points $A$ and $B$, 
representing two images of the same tachyon.
Source $A$ is eventually absorbed at $z'=0$, abruptly diminishing the wave amplitude (as will be 
shown later, by half). 

\begin{figure}[h]
\begin{center}
\includegraphics*[width=0.5\textwidth]{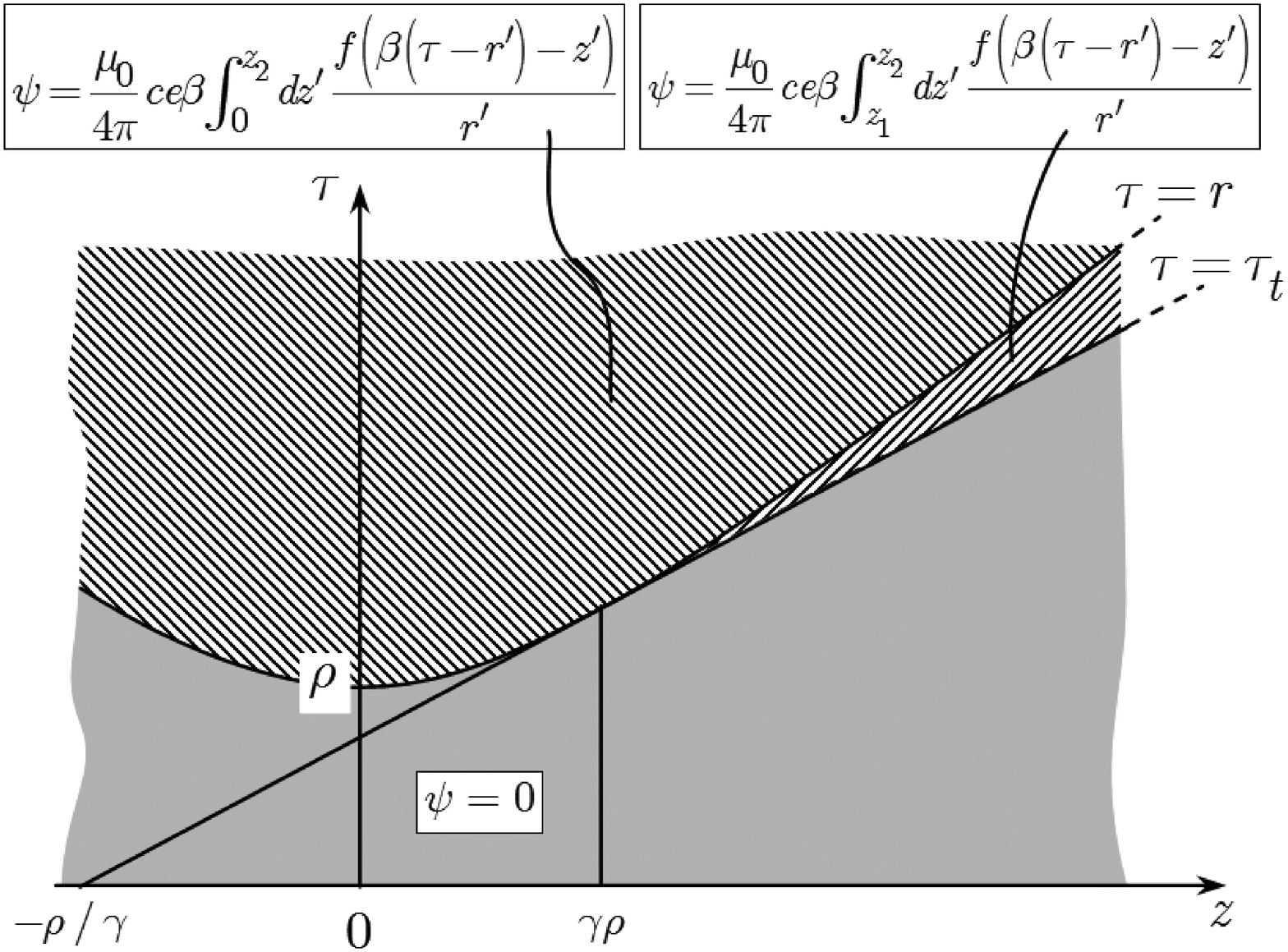}
\end{center}
\caption{Areas of case representation of $\psi \left(\tau, \rho, z \right)$.}
\label{Fig_case_areas_z_tau}
\end{figure}

The $z,\tau $ plane map of the three areas differing in the solution 
representation is shown in Fig.~\ref{Fig_case_areas_z_tau}.
%
%
%
Aiming at construction of the propagation-invariant waves, it is 
worthwhile to express the wavefunction in terms of the propagation variable $\xi = \beta \tau - z$. 
In contrast to the pure steady-state solutions, representation of wavefront-limited 
causal waves does not exclude the time dependence, and passing from 
$\tau, \rho, z$ to $\tau, \rho , \xi$ \textit{transforms}:

(i) the non-zero wave condition $\tau > \tau_t$ \textit{into} 
$\rho <\gamma \xi $, a causal analog of the condition 
$\rho <\gamma \left| \zeta \right|=\gamma \left| \xi \right|$, 
reported in (\cite{RecamiZamboniRachedDartora2004}, Eq.~(8))
for the steady-state electromagnetic field of a charged tachyon ---
in agreement with predictions expound by Recami et al. on the basis of the 
extended theory of special relativity (see \cite{Barut_1982}, especially Fig.~4, as well as
earlier works on superluminal Lorentz transformations
\cite{RecamiMaccarrone_ProblIm1_1980,CaldirolaEtAl_ProblIm2_1980});

(ii) the case-limiting condition $\tau < r(\tau, \rho, \xi)=\sqrt{{{\rho }^{2}}+{{\left( \beta \tau -\xi  \right)}^{2}}}$,
 \textit{into} ${{\tau }_{1}}<\tau <{{\tau }_{2}}$, where
\begin{equation} \label{tau12}
\tau_{1,2} 
= \tau - z_{2,1}/\beta 
= \gamma^2\left( \beta \xi \mp \sqrt{{{\xi }^{2}}-
  \left( {{\beta }^{2}}-1 \right){{\rho }^{2}}} \right)
\end{equation}
are two roots of the equation
\begin{equation} \label{Eq_for_tau12}
{{\tau }_{1,2}}-r(\tau_{1,2}, \rho, \xi)=0;
\end{equation}

(iii) case formulas (\ref{PsiOrdinaryCases}), (\ref{PsiExtraordinaryCases}) \textit{into} \\
$\texttt{Case }-\infty <\xi <\rho /\gamma \texttt{ or (}
\rho /\gamma <\xi <\infty \texttt{ and }-\infty <\tau <{{\tau }_{1}}\texttt{)}$
\begin{equation} \label{psi_CaseA}
{\psi}\left(\tau, \rho, \xi \right) =0,
\end{equation}
$\texttt{Otherwise }$ $\texttt{Case }{{\tau }_{1}}<\tau <{{\tau }_{2}}$ 
\begin{equation} \label{psi_CaseB}
{\psi}\left(\tau, \rho, \xi \right) =
     \frac{{{\mu }_{0}}}{4\pi }ce{{\beta }^{2}}
     \int_{\tau_1}^{\tau } {
       d {\tau }'\frac{f\left( \beta \left[ {\tau }' - r \left(\tau', \rho, \xi \right) \right] \right)}
       {r \left(\tau', \rho, \xi \right)}},
\end{equation}
$\texttt{Case }{{\tau }_{2}}<\tau <\infty$\\
\begin{equation} \label{psi_CaseC}
{\psi}\left(\tau, \rho, \xi \right) =
     \frac{{{\mu }_{0}}}{4\pi }ce{{\beta }^{2}}
     \int_{\tau_1}^{\tau_2} {
       d {\tau }' \frac{f\left( \beta \left[ {\tau }'-r \left(\tau', \rho, \xi \right) \right] \right)}
       {r \left(\tau', \rho, \xi \right)}};
\end{equation}
%
%
%

\begin{figure}[h]
\begin{center}
\includegraphics*[width=0.5\textwidth]{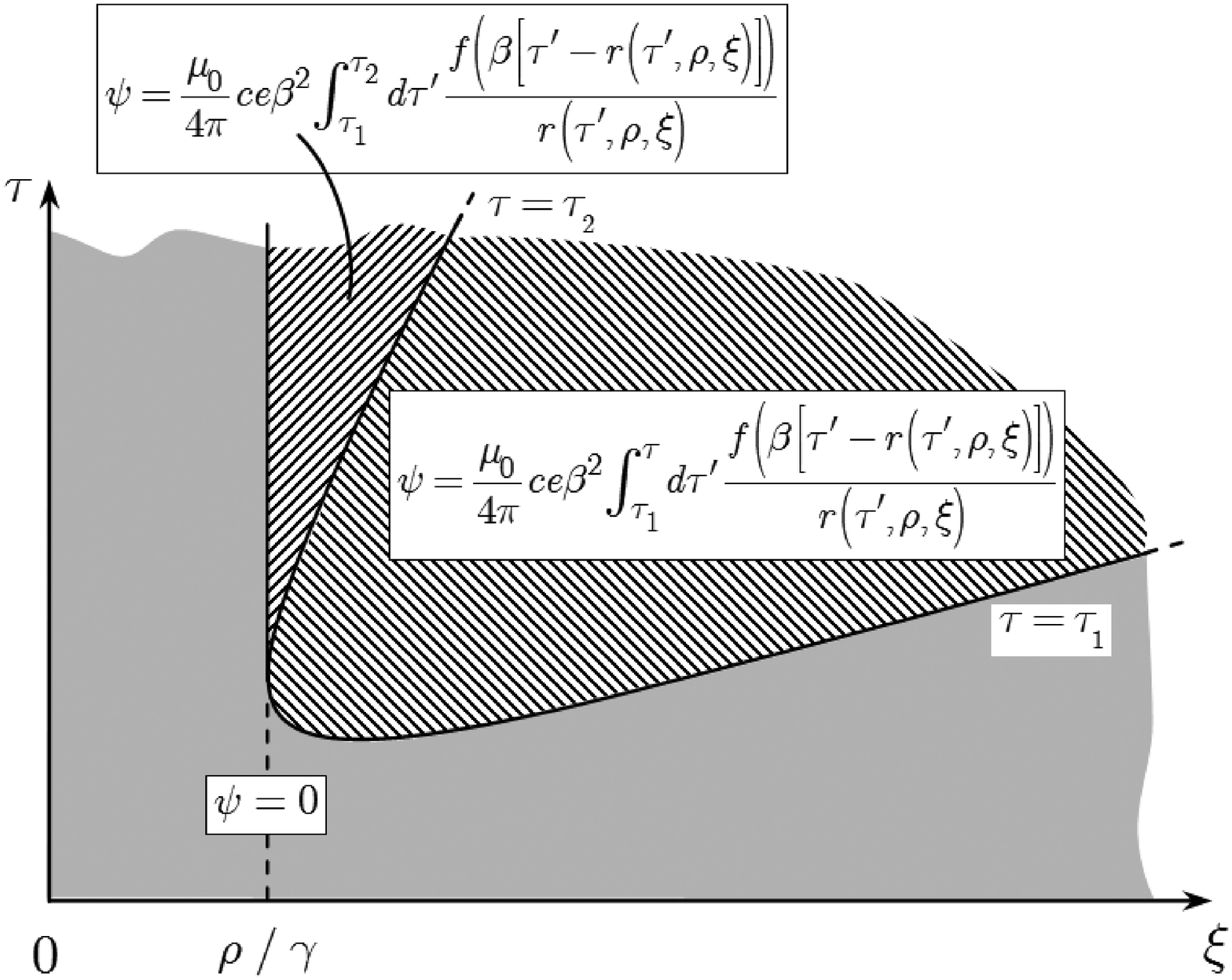}
\end{center}
\caption{Areas of case representation of $\psi \left( \tau, \rho, \xi \right)$.}
\label{Fig_case_areas_xi_tau}
\end{figure}

(iv) the case map of Fig.~\ref{Fig_case_areas_z_tau} ($z, \tau$ plane) \textit{into} the 
case map represented in Fig.~\ref{Fig_case_areas_xi_tau} ($\xi, \tau$ plane).
%
%
%

\begin{figure}[ht]
\begin{center}
\includegraphics*[width=0.8\textwidth]{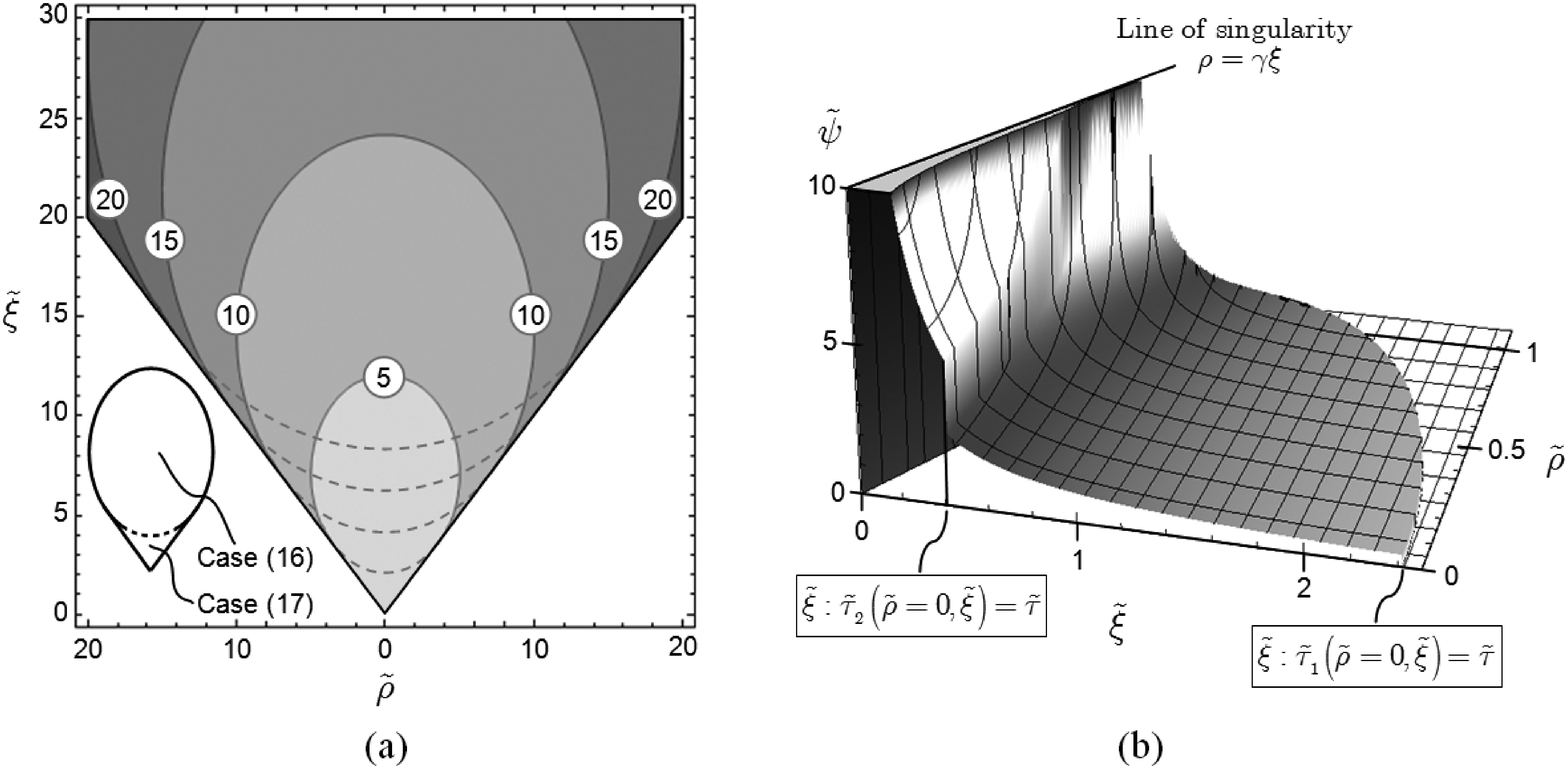}
\end{center}
\caption{(a) Contour plots illustrating the shape of the wavefunction support (solid isolines 
$\tilde{\tau }_1 = \tilde{\tau }$) and the boundary between cases (\ref{psi_CaseB}) and 
(\ref{psi_CaseC}) (dashed isolines ${\tilde{\tau }}_{2} = \tilde{\tau }$); (b) a snapshot 
of $\tilde{\psi}$, clipped by $\tilde{\psi} = 10$ in the vicinity of its singularities,
taken at $\tilde{\tau } = 1$.}
\label{Fig_wfunction_supp}
\end{figure}

Contour plot, representing the wave propagation via a characteristic space-time scale
$\lambda_0$ and the dimensionless quantities
$\tilde{\tau }=\tau /{{\lambda }_{0}}$,
${{\tilde{\tau }}_{1,2}}={{\tau }_{1,2}}/{{\lambda }_{0}}$, 
$\tilde{\rho }=\rho /{{\lambda }_{0}}$, 
$\tilde{\xi }=\xi /{{\lambda }_{0}}$,
is given in Fig.~\ref{Fig_wfunction_supp}(a) 
($\beta =\sqrt{2}$, as in numerical illustrations of Ref.~\cite{RecamiZamboniRachedDartora2004}).
For each fixed moment of time $\tilde{\tau }$ the solid isoline 
$\tilde{\tau }_1(\tilde{\rho}, \tilde{\xi}) = \tilde{\tau }$ 
defines the boundary of the wavefunction support (area $\tilde{\tau}_1 < \tilde{\tau }$ 
corresponding to $\psi \ne 0$) while the dashed isoline 
${{\tilde{\tau }}_{2}}(\tilde{\rho}, \tilde{\xi})=\tilde{\tau }$ 
traces the boundary between cases (\ref{psi_CaseB}) and (\ref{psi_CaseC}). 
The emanated wave has an expanding droplet-shaped support, defined exclusively by 
the parameters $\rho, \xi, \tau$, and $\beta$.
Fig.~\ref{Fig_wfunction_supp}(b) represents the structure of this droplet-shaped wave
for $f\left( \xi  \right)=\delta \left( \xi  \right)$; the wavefunction is represented
in dimensionless units via normalization 
$ \tilde{\psi}={\psi}/{\psi_0}, \psi_0 =\frac{\mu_0 ce}{2\pi\lambda_0}$.

%
%
%
%
\section{Droplet-shaped waves as causal counterparts of the X-shaped waves}
\label{sec:4}

\begin{figure}[t]
\begin{center}
\includegraphics*[width=1.0\textwidth]{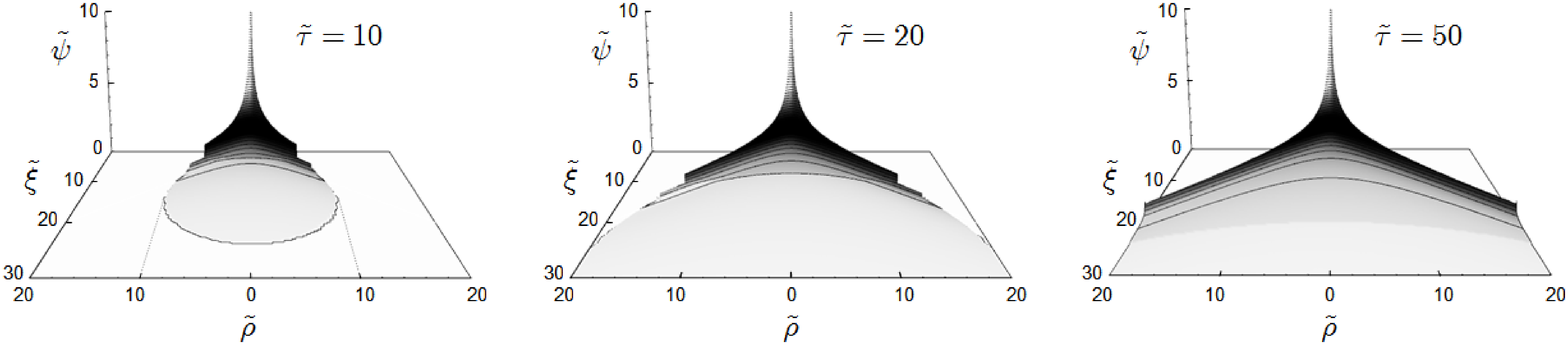}
\end{center}
\caption{Dynamics of the droplet-shaped wave propagation (the points on the vertex and the cone surface, in which $\tilde{\psi}$ diverges, are omitted).}
\label{Fig_drop_wave_dynam}
\end{figure}

Using the above approximation of an infinitely short source current pulse 
$f\left( \xi  \right)=\delta \left( \xi  \right)$, 
which in the case discussed in \cite{RecamiZamboniRachedDartora2004} led to the model of X wave 
generation by a charged tachyon, one readily arrives at the solution that in the area 
$\xi >\rho /\gamma $ and $\tau >{{\tau }_{1}}$ ($\psi \ne 0$) reads
\begin{equation} \label{PsiCasesForDeltaPulse}
{\psi}  =  \left\{ 
\begin{array}{ll}
 \frac{{{\mu }_{0}}}{4\pi }ce\beta 
 \int_{{{\tau }_{1}}}^{\tau } {
   d {\tau }' \frac{\delta \left( {\tau }'-r \left(\tau', \rho, \xi \right) \right)}{{{\tau }'}}
 } & \; \tau_1 < \tau < \tau_2  \\
 \frac{{{\mu }_{0}}}{4\pi }ce\beta 
 \int_{{{\tau }_{1}}}^{{{\tau }_{2}}} {
   d {\tau }' \frac{\delta \left( {\tau }'-r \left(\tau', \rho, \xi \right) \right)}{{{\tau }'}}
 } & \; \tau_2 < \tau <\infty.
\end{array} 
\right.
\end{equation}
The equation for the delta-function roots coincides with (\ref{Eq_for_tau12}), 
so these roots, ${\tau }'={{\tau }_{1,2}}$, are defined by formula (\ref{tau12}). 
Using (\ref{DeltaFuncWithSimple0}) and passing to the dimensionless parameters finally reduce (\ref{PsiCasesForDeltaPulse}) to the formula
\begin{equation} \label{PsiForDeltaDimless}
\tilde{\psi}\left(\tilde{\tau}, \tilde{\rho}, \tilde{\xi}\right)=\left\{ 
\begin{array}{ll}
   \frac{1}{2}\frac{\beta }{\sqrt{{{{\tilde{\xi }}}^{2}}-\left( {{\beta }^{2}}-1 \right){{{\tilde{\rho }}}^{2}}}} 
   & \; {{{\tilde{\tau }}}_{1}}<\tilde{\tau }<{{{\tilde{\tau }}}_{2}}  \\
   \frac{\beta }{\sqrt{{{{\tilde{\xi }}}^{2}}-\left( {{\beta }^{2}}-1 \right){{{\tilde{\rho }}}^{2}}}}
   & \; {{{\tilde{\tau }}}_{2}}<\tilde{\tau }<\infty.
\end{array} \right.
\end{equation}
One can check by direct calculation that the second case of (\ref{PsiForDeltaDimless})
corresponds to solution of the inhomogeneous wave equation of Ref.~\cite{RecamiZamboniRachedDartora2004}, 
describing the steady-state X wave produced by a charged tachyon (delta-pulse source).
Introduction of the initial moment of particle generation results in launching of the same 
propagation-invariant waveform, which however is restricted by the droplet-shaped support
illustrated in Fig.~\ref{Fig_wfunction_supp}(a). As the time increases, this waveform 
expands in both 
$\tilde{\rho }$ and $\tilde{\xi }$ directions, tending to the charged-tachyon X-shaped field, 
as illustrated in Fig.~\ref{Fig_drop_wave_dynam}.
%
%
%
%
For ${{\tilde{\tau }}_{1}}<\tilde{\tau }<{{\tilde{\tau }}_{2}}$ the singularity 
corresponding to ${{\tilde{\tau }}_{2}}$ (in Fig.~\ref{Fig_zp_taup_sequence}, to $z_1$) 
resides outside the integration segment $\Gamma_i$, diminishing $\psi$ by half.

Corresponding electromagnetic field vectors 
${\bf E} =  -c\left(\nabla A^{(0)}+\partial{\bf A }/\partial \tau\right)$ and 
${\bf{B}} = \nabla \times {\bf{A}}$ (of which only $E_\rho$, $E_z$, 
and $B_\varphi$ components are nonzero), can readily be found using Eqs. (\ref{FormulaForElCharge})
and (\ref{FormulaForAo}). In particular, for the observation times 
$\tilde{\tau }>{{\tilde{\tau }}_{2}}$ (a steady-state wave zone located 
outside the singularities arising from the potential discontinuities on the case 
delimiting boundaries of Fig.~\ref{Fig_case_areas_xi_tau}) 
the magnetic field (normalized by $\psi_0 \lambda_0^{-1}$) 
is characterized by
${\tilde B_\varphi } 
=  - {\partial \tilde \psi }/{\partial \tilde \rho } 
= -\tilde \rho \beta 
\left( {{\beta ^2} - 1} \right)
\left[ 
{{\tilde \xi }^2} - \left( {{\beta ^2} - 1} \right){{\tilde \rho }^2}
\right]^{ - 3/2}$. As in the case described in \cite{RecamiZamboniRachedDartora2004}, Sec.~III, 
it remains the only component that does not vanish when $\beta \rightarrow \infty$, 
revealing a "magnetic monopole" behavior. As put forward in \cite{RecamiReview_RivistaNuovoCim_1986,Barut_1982,RecamiMaccarrone_ProblIm1_1980,CaldirolaEtAl_ProblIm2_1980},
the reference frame in which the particle velocity tends to infinity plays 
for tachyons the same role as the rest frame for ordinary particles (bradyons), 
and there exists a duality between subliminal electric charges and superluminal 
magnetic monopoles. So, for $\beta \rightarrow \infty$ one might expect
the magnetic field to have a structure akin to that of the electric field of a charged 
particle. 
Notably, earlier applications of 
the proposed technique to the waves emanated by subluminal sources (see, for instance, 
\cite{UtkinBookCh2011,BorUtk_JPD_1995}) result in peak- or ball-like shapes akin to 
the subliminal wave bullets obtained in Secs. II-IV of Ref.~\cite{ZamboniRecami_PRA_2008}.
While electromagnetic fields of different line currents propagating at luminal speed 
present singularity at one "point of accumulation", $\xi \equiv \tau-z=0, \rho=0$ (see, e.g., Fig.~1
of \cite{BorUtk_JMP_1994}), for the superluminal delta current the area of singularity
spreads along the conical surface $\xi - \rho/\gamma = 0$, representing, according to
\cite{RecamiReview_RivistaNuovoCim_1986,Barut_1982} 
an initially point-like structure (perceived by superluminal observers) highly distorted
by the superluminal Lorentz transformation. 
In toto, the results obtained support the general
idea about the shape of superluminal particles and distribution of the field associated 
with superluminal charges: "while the simplest subluminal object is obviously a sphere 
or, in the limit, a space point, the simplest superluminal object is on the contrary an 
X-shaped pulse" \cite{ZamboniRecami_PRA_2008}. 

\section{Conclusion}
\label{sec:5}

The present work introduces a new type of the propagation-invariant, localized droplet-shaped waves that, 
being solutions of the inhomogeneous wave equation and satisfying zero initial conditions, admit causal 
generation by real pulsed superluminal sources widely discussed in the literature (see in 
particular the monograph \cite{RecamiBook1978} and more recent review \cite{RecamiReview2001} by Recami). 
Although the presented analysis is limited to the most illustrative case of $\delta$-pulse current, the 
general integral solution (\ref{psi_CaseA})-(\ref{psi_CaseC}) enables the propagation-invariant 
localized droplet-shaped waves emanated 
by various line sources to be investigated both analytically and numerically using the ${z}',{\tau }'$ 
diagrams similar to one of Fig.~\ref{Fig_zp_taup_sequence}. 
Various possible ans\"{a}tze are discussed in \cite{UtkinBookCh2011} for free space and 
\cite{Utkin_WM_2012} for waveguides.

\section*{Acknowledgements}
The author is grateful to Michel Zamboni-Rached and Erasmo Recami for several useful discussions that 
significantly improved the introductory part of the article as well as for their kind cooperation that 
stimulated publication of this study. The research was partially based on work supported 
by the International Science Foundation under Grant M3H000 
{\it Generation and Propagation of Localized Waves}.

\end{document}